\documentclass[aps,prd,preprint,tightenlines,superscriptaddress,showpacs,byrevtex]{revtex4}
\usepackage{graphicx}
\usepackage{epsfig}
\usepackage{dcolumn}
\newcommand{\y}{Y(2175)}

\newcommand{\BR}{{\cal B}}
\newcommand{\jpc}{J^{PC}}

\newcommand{\EE}{e^+e^-}

\newcommand{\pp}{\pi^+\pi^-}
\newcommand{\fzero}{f_0(980)}

\newcommand{\beq}{\begin{equation}}
\newcommand{\eeq}{\end{equation}}
\newcommand{\beqy}{\begin{eqnarray}}
\newcommand{\eeqy}{\end{eqnarray}}
\newcommand{\bitm}{\begin{itemize}}
\newcommand{\eitm}{\end{itemize}}

\begin{document}

%************************************************************
\preprint{\vbox{ \hbox{   }
                 \hbox {  }
                        \hbox{BIHEP-EP-2009-002}
                         }}
\title{\quad\\[1.0cm]
Combined fit to BaBar and Belle Data on $\EE \to \phi\pp$ and
$\phi \fzero$}

\affiliation{Institute of High Energy Physics, Chinese Academy of
Sciences, Beijing 100049, China}

\affiliation{University of Hawaii, Honolulu, HI 96822, USA}

\author{C.~P.~Shen}
\email{shencp@phys.hawaii.edu} \affiliation{University of Hawaii,
Honolulu, HI 96822, USA}
\author{C.~Z.~Yuan}
\email{yuancz@ihep.ac.cn} \affiliation{Institute of High Energy
Physics, Chinese Academy of Sciences, Beijing 100049, China}

\date{\today}

\begin{abstract}

A combined fit is performed to the BaBar and Belle measurements of
the $\EE \to \phi \pp$ and $\phi \fzero$ cross sections for
center-of-mass energy between threshold and 3.0~GeV. The resonance
parameters of the $\phi(1680)$ and $\y$ are determined. The mass
is $(1681^{+10}_{-12})$~MeV/$c^2$ and the width is
$(221^{+34}_{-24})$~MeV/$c^2$ for the $\phi(1680)$, and the mass
is $(2117^{+59}_{-49})$~MeV/$c^2$ and the width is
$(164^{+69}_{-80})$~MeV/$c^2$ for the $\y$. These information will
shed light on the understanding of the nature of the excited
$\phi$ and $Y$ states observed in $\EE$ annihilation.

\end{abstract}


\maketitle

%%%%%%%%%%%%%%% Introduction %%%%%%%%%%%%%%%%%%%%%%%%%%%%%%%%%%%%%%%%
\section {Introduction}

Although vector mesons are produced copiously in $\EE$
annihilation, the resonance parameters of them are not well
measured~\cite{PDG}. In the $s\bar{s}$ sector, in a study of
initial-state radiation (ISR) events of the type, $\EE \to
\gamma_{\rm ISR} \phi \pp$, the BaBar and Belle Collaborations
observed two clear structures near $\sqrt{s}=1.68$~GeV and
2.175~GeV, the latter was produced dominantly via a $\phi \fzero$
intermediate state, and was dubbed the
$\y$~\cite{babar2175,belle-phipp}. The $\phi(1680)$ was measured
by the Belle Collaboration with its mass and width to be $1689\pm
7\pm 10$~MeV/$c^2$ and $211\pm 14\pm 19$~MeV/$c^2$,
respectively~\cite{belle-phipp}. For the $\y$, its mass and width
measured by the BaBar Collaboration are $2175 \pm 10 \pm
15$~MeV/$c^2$ and $58\pm 16\pm 20$~MeV/$c^2$, respectively, by
fitting the $\phi \fzero$ cross section; while with more
luminosity, the Belle experiment found that its mass and width are
$2163\pm 32$~MeV/$c^2$ and $125\pm 40$~MeV/$c^2$, respectively.
Examining closely the fitting to the $\phi \fzero$ cross section,
we found that the fitted $\y$ resonance parameters are sensitive
to the background shape. So it is necessary to give a more
sophisticated estimation of the resonance parameters of the $\y$
for a better understanding of its nature. This goal can be reached
by performing a combined fit to the $e^+ e^- \to \phi\pi^+\pi^-$
and $e^+ e^- \to \phi \fzero$ cross sections measured by the BaBar
and Belle experiments since all the data are available via Durham
database~\cite{babar2175,belle-phipp}.

Since the $\y$ resonance is produced via ISR in $\EE$ collisions,
its $\jpc=1^{--}$. The $\y$ was firstly suspected to be an
$s$-quark partner of the $Y(4260)$~\cite{babay4260,bn978} since
both are produced in $\EE$ annihilation and exhibit similar decay
patterns. On the other hand, a number of different interpretations
have been proposed, which include: an $s\bar{s}g$
hybrid~\cite{ssg}; a $2^3D_1$ $s\bar{s}$ state~\cite{2D} with a
width predicted to be in the range 120-210~$\hbox{MeV}/c^2$; a
tetraquark state~\cite{4s,chenhx,dren}; a $\Lambda \bar{\Lambda}$
bound state~\cite{eberhard}; a conventional excited $\phi$
meson~\cite{susana}; or a structure produced by final state
interactions~\cite{27mev,alv}. The possibility that the $\y$ is a
$3^3S_1$ $s\bar{s}$ state is disfavored by the rather large
predicted width ($\Gamma \sim 380~\hbox{MeV}/c^2$)~\cite{3s}. A
review~\cite{zhusl} discusses the basic problem of the large
expected decay widths into two mesons, which are in contrast to
experimental observations.

In this paper, we try to perform all the possible fits to the $e^+
e^- \to \phi\pi^+\pi^-$ and $e^+ e^- \to \phi \fzero$ cross
sections measured by the BaBar and Belle experiments to obtain a
better knowledge of the resonance parameters of the $\phi(1680)$
and $\y$. We also study the possible production of the structure
at around 2.4~GeV/$c^2$ in both $\phi \pp$ and $\phi \fzero$
modes.

%%%%%%%%%%%%%% the data %%%%%%%%%%%%%%%%%%%%%%%%%%%%%%%%%%%%%%%%%
\section {The data}

Both the BaBar and the Belle experiments reported cross sections
of $e^+ e^- \to \phi\pi^+\pi^-$ and $e^+ e^- \to \phi \fzero$ for
center-of-mass energy ranges from threshold to about
3.0~GeV~\cite{babar2175,belle-phipp}. The integrated luminosity of
the BaBar data sample is 232~fb$^{-1}$ while that of the Belle
data sample is 673~fb$^{-1}$, with $\sim90\%$ of the data
collected at the $\Upsilon(4S)$ resonance ($\sqrt{s}=10.58$~GeV),
while the rest were taken off the $\Upsilon(4S)$ peak.

\begin{figure*}[htbp]
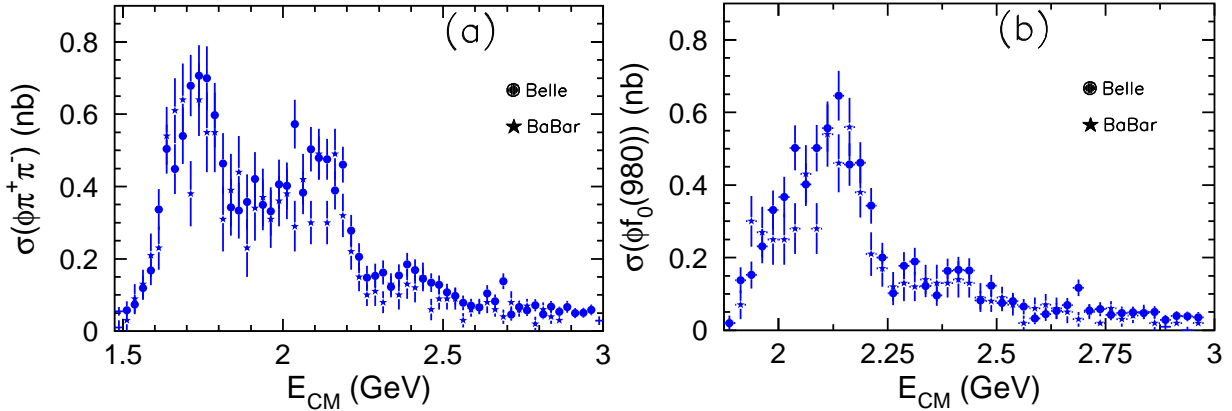

\centerline{\psfig{file=fig1a.epsi,width=8cm}
\psfig{file=fig1b.epsi,width=8cm} } \caption{The cross sections of
$e^+ e^- \to \phi\pi^+\pi^-$ (a) and $\phi \fzero$ (b) measured at
BaBar (stars with error bars) and Belle (dots with error bars). }
\label{cross}
\end{figure*}

Figure~\ref{cross} shows the data, good agreement between BaBar
and Belle results is observed within errors, and the two
structures (the $\phi(1680)$ and $\y$) are evident in $e^+ e^- \to
\phi\pi^+\pi^-$ mode, and one structure (the $\y$) is evident in
$e^+ e^- \to \phi \fzero$ mode. Also we notice that there are
clusters of events near 2.4~GeV/$c^2$ in both modes.

%%%%%%%%%%%%%%%%% phi pi+ pi-  cross section %%%%%%%%%%%%%%%%%%%%%%%%%
\section {The fit to $\EE\to \phi\pp$ cross section}

In order to obtain the resonance parameters of the $\phi(1680)$
and $\y$, we fit the $\phi \pp$ cross section distribution using a
least square method with MINUIT in the CERN Program
Library~\cite{minuit}. The Breit-Wigner ($BW$) form of a signal
resonance used in $\EE\to\phi \pp$ analysis is
$$
BW(\sqrt{s})=\sqrt{\frac{M^2}{s}}\frac{\sqrt{12\pi\Gamma_{\EE}\BR(R\to
f)\Gamma_{{\rm tot}}}}{s-M^2+iM\Gamma_{{\rm
tot}}}\sqrt{\frac{PS(\sqrt{s})}{PS(M)}}
$$
where $M$ is the mass of the resonance, $\Gamma_{{\rm tot}}$ and
$\Gamma_{\EE}$ are the total width and partial width to $\EE$
respectively, $\BR(R\to f)$ is the branching fraction of $R$
decays into final state $f$, and $PS(\sqrt{s})$ is the three-body
decay phase space factor. We fit the Belle data on $\phi\pp$
between threshold and 3.0~GeV (61 data points) and the BaBar data
in the similar energy range (56 data points) simultaneously.

\begin{figure*}[htbp]
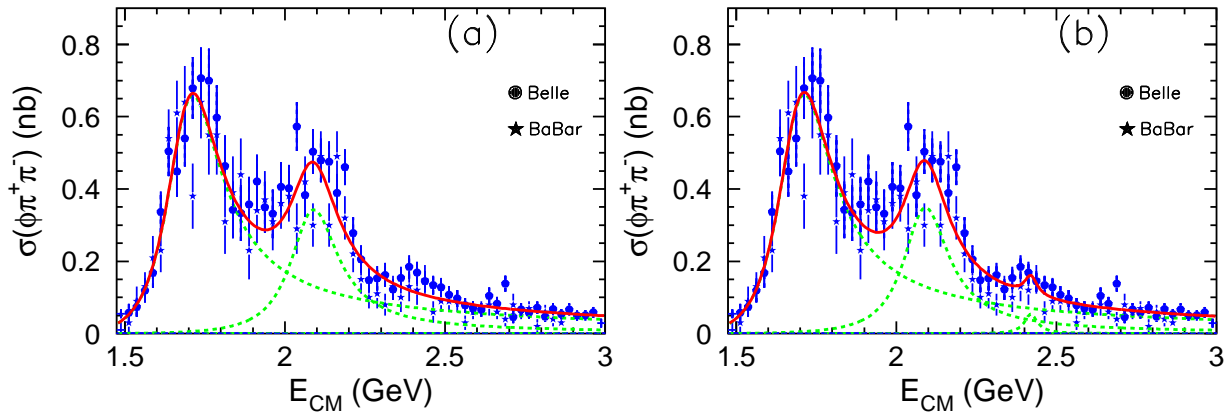

\centerline{\psfig{file=fig2a.epsi,width=8cm}
\psfig{file=fig2b.epsi,width=8cm} } \caption{Fit to $\EE\to \phi
\pp$ cross sections with two incoherent $BW$ functions (a), and
with three incoherent $BW$ functions (b), respectively. The curves
show the projections from the best fit and the contribution from
each component.} \label{phipp-nocoh}
\end{figure*}

Since the $\phi(1680)$ decays into $\phi \pp$ while the $\y$
decays dominantly into $\phi \fzero$, we use two incoherent $BW$
functions in the fit first, one for the $\phi(1680)$ and the other
for the $\y$. The fit result is shown in
Fig.~\ref{phipp-nocoh}(a), with a goodness-of-the-fit of
$\chi^2/ndf=170/111$, corresponding to a confidence level (C.L.)
of 0.03\%. The statistical significance of each resonance is
greater than 10$\sigma$. From the fit we obtain the following
resonance parameters of the $\phi(1680)$: $M=(1685\pm
5)$~MeV/$c^2$, $\Gamma_{\rm tot}=(208\pm 11)$~MeV/$c^2$ and
$\BR(\phi \pp)\times \Gamma_{\EE}=(24.6\pm1.2)~ \hbox{eV}/c^2$,
while those of the $\y$ are $M=(2080\pm 12)$~MeV/$c^2$,
$\Gamma_{\rm tot}=(182\pm 20)$~MeV/$c^2$ and $\BR(\phi \pp)\times
\Gamma_{\EE} = (18.1\pm 1.8)~\hbox{eV}/c^2$, where the errors are
statistical only. A fit with an additional nonresonant component
does not improve the fit quality, and the contribution of the
nonresonant term is negligibly small and can be neglected.

Since there are some events accumulating at about 2.4~GeV, we also
perform a fit with an additional incoherent $BW$ function for this
structure. The fitted parameters of this structure are $M=(2419\pm
25)$~MeV/$c^2$, $\Gamma_{\rm tot}=(49\pm 42)$~MeV/$c^2$ and
$\BR(\phi \pp)\times \Gamma_{\EE}=(1.06\pm 0.58)~\hbox{eV}/c^2$
with a goodness-of-the-fit of $\chi^2/ndf=163/108$, corresponding
to a C.L. of 0.05\%. The statistical significance of the structure
is $1.8\sigma$ as determined from the change in the $\chi^2$ value
compared to the two-incoherent-resonance fit and the difference in
the number of degrees of freedom. The fitted values of the
$\phi(1680)$ and $\y$ resonance parameters only change a little
compared to the two-incoherent-resonance fit, and the difference
is quoted as one source of systematic errors. The fit result is
shown in Fig.~\ref{phipp-nocoh}(b).

\begin{figure}[htbp]
\centerline{\psfig{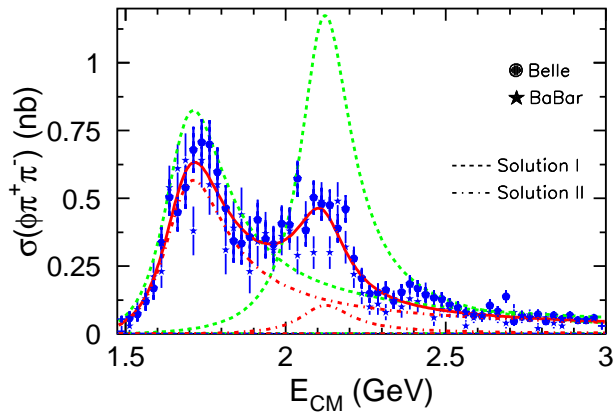} } \caption{ The
results of the fit to $\EE \to \phi \pp$ data from Belle and
BaBar. The curve show the the projection from the best fit with
two coherent $BW$s. The interference between the two amplitudes is
not shown. The two dashed curves at each peak show the two
solutions (see text), one is for the destructive solution
(Solution I), the other is for the constructive solution (Solution
II).} \label{phipp-twocoh}
\end{figure}

We also fit with two coherent $BW$ functions.
Figure~\ref{phipp-twocoh} shows the fit results. There are two
solutions with equally good fit quality, with a
goodness-of-the-fit of $\chi^2/ndf=158/110$, corresponding to a
C.L. of 0.19\%. The masses and the widths of the two resonances
are identical but the partial widths to $\EE$ and relative phases
are different in these two solutions, as shown in
Table~\ref{phipp-twosolution}. The resonance parameters of the
$\phi(1680)$ are $M=(1677\pm 6)$~MeV/$c^2$ and $\Gamma_{\rm
tot}=(233\pm 16)$~MeV/$c^2$, and those of the $\y$ are $M=(2112\pm
16)$~MeV/$c^2$ and $\Gamma_{\rm tot}=(196\pm 27)$~MeV/$c^2$.

\begin{table}[htbp]
\caption{Fit results to the combined BaBar and Belle data on $\EE
\to \phi \pp$ with two coherent $BW$ functions. The errors are
statistical only. $M$, $\Gamma_{{\rm tot}}$, and $\BR(\phi
\pp)\times \Gamma_{\EE}$ are the mass (in~MeV/$c^2$), total width
(in~MeV/$c^2$), and product of the branching fraction to $\phi
\pp$ and the $\EE$ partial width (in~eV/$c^2$), respectively.
$\phi$ is the relative phase (in degree).}
\label{phipp-twosolution}
\begin{center}
\begin{tabular}{c c c  c }
\hline \hline
  Parameters &  & Solution I & Solution II \\\hline
 $M(\phi(1680))$& & \multicolumn{2}{c}{$1677\pm6$} \\
 $\Gamma_{{\rm tot}}(\phi(1680))$& & \multicolumn{2}{c}{$233\pm16$}  \\
 $\BR(\phi \pp)\times \Gamma_{\EE}(\phi(1680)) $ & & $33.4\pm1.3$ & $23.0\pm1.5$  \\
 $M(\y)$ & & \multicolumn{2}{c}{$2112\pm16$} \\
  $\Gamma_{{\rm tot}}(\y)$& &   \multicolumn{2}{c}{$196\pm27$}\\
 $\BR(\phi \pp)\times
\Gamma_{\EE}(\y)$ & & $68.9\pm7.0$ & $6.2\pm1.1$\\
 $\phi$ &  & $-122\pm2$ & $90\pm8$ \\
 \hline \hline
\end{tabular}
\end{center}
\end{table}

\begin{figure*}[htbp]
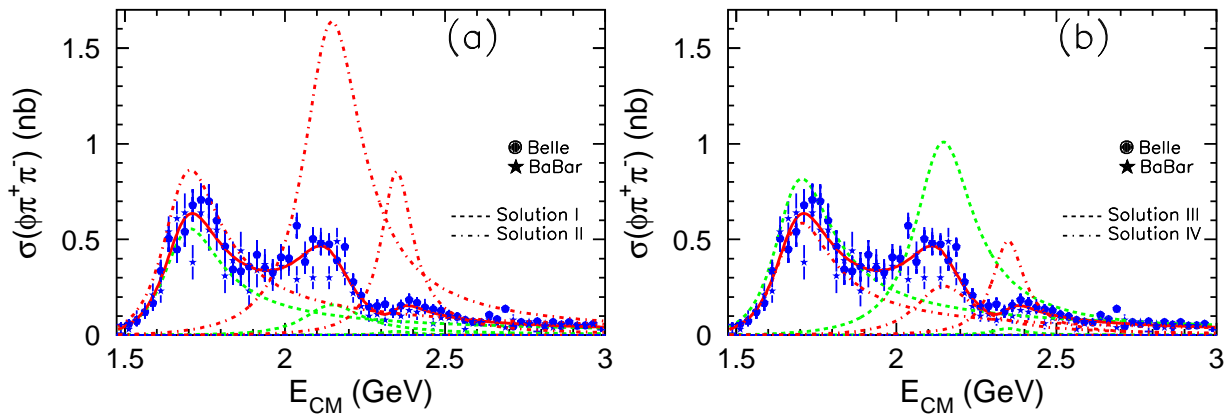

\centerline{\psfig{file=fig4a.epsi,width=8cm}
\psfig{file=fig4b.epsi,width=8cm} } \caption{The results of the
fit to $\EE \to \phi \pp$ data from Belle and BaBar with three
coherent $BW$ functions. The curves show the projections from the
best fit with three coherent $BW$s. There are four solutions. }
\label{belle-three-cohe}
\end{figure*}

Similarly we also try to perform a fit with an additional coherent
$BW$ function at around 2.4~GeV/$c^2$. The result is shown in
Fig.~\ref{belle-three-cohe}. There are four solutions. The values
of the fitted results for these four solutions are shown in
Table~\ref{phipp-foursolution}, with a goodness-of-the-fit of
$\chi^2/ndf=143/106$.  The fitted values of the masses and widths
of the $\phi(1680)$ and $\y$ are consistent with the above results
within 2$\sigma$, and the statistical significance of the
structure at $2.4~\hbox{GeV}/c^2$ is estimated to be $2.8\sigma$
as determined from the change in the $\chi^2$ value compared to
the two-coherent-resonance fit and the difference in the number of
degrees of freedom.

\begin{table*}[htbp]
\caption{Fit results to the combined BaBar and Belle data on $\EE
\to \phi \pp$ with three $BW$ functions that interfere with each
other. The errors are statistical only. $M$, $\Gamma_{{\rm tot}}$,
and $\BR(\phi \pp)\times \Gamma_{\EE}$ are the mass
(in~MeV/$c^2$), total width (in~MeV/$c^2$), and product of the
branching fraction to $\phi \pp$ and the $\EE$ partial width
(in~eV/$c^2$), respectively. $\phi$ is the relative phase (in
degree).} \label{phipp-foursolution}
\begin{center}
\begin{tabular}{c  c  c c c}
\hline \hline
  Parameters & Solution I & Solution II & Solution III & Solution IV \\\hline
$M(\phi(1680))$ & \multicolumn{4}{c}{$1673\pm6$} \\
  $\Gamma_{{\rm tot}}(\phi(1680))$&   \multicolumn{4}{c}{$218\pm 13$}\\
 $\BR(\phi \pp)\times
\Gamma_{\EE}(\phi(1680))$ & $21.2\pm1.1$ & $32.9\pm1.5$ & $31.2\pm1.5$ & $22.3\pm1.1$\\
 $M(\y)$ & \multicolumn{4}{c}{$2133\pm24$} \\
  $\Gamma_{{\rm tot}}(\y)$&   \multicolumn{4}{c}{$221\pm38$}\\
 $\BR(\phi \pp)\times
\Gamma_{\EE}(\y)$ & $10.7\pm1.8$ & $110.4\pm16.0$ & $68.0\pm5.5$ & $17.3\pm4.6$\\
 $\phi_1$ & $110\pm5$ & $-111\pm5$ & $-126\pm3$ & $125\pm9$\\
 $M(X(2400))$ & \multicolumn{4}{c}{$2346\pm26$} \\
  $\Gamma_{{\rm tot}}(X(2400))$&   \multicolumn{4}{c}{$121\pm35$}\\
 $\BR(\phi \pp)\times
\Gamma_{\EE}(X(2400))$ & $0.90\pm0.37$ & $38.6\pm15.7$ & $1.6\pm0.8$ & $22.3\pm5.9$\\
 $\phi_2$ & $28\pm23$ & $159\pm5$ & $-127\pm28$ & $-46\pm6$\\
 \hline \hline
\end{tabular}
\end{center}
\end{table*}

%%%%%%%%%%%%%%%%%%%%%%%%%%%%%%%%%%%%%%%%%%%%%%%%%%%%%%%%%%%%%%%%%%%%
\section {The fit to $\EE\to \phi\fzero$ cross section}

We fit the Belle data on $\phi\fzero$ between threshold and
3.0~GeV (44 data points) and the BaBar data in the same energy
range (44 data points) simultaneously. For the nonresonant
component amplitude, we use the same formula used by the BaBar
Collaboration~\cite{babar2175} as the following:
$$
\sigma_{nr}=P(\mu) \times |A_{nr}(\mu)|^2,
$$
$$
A_{nr}(\mu)=\sqrt{\sigma_0}\times (1-e^{-(\mu/a_1)^4}) \times (1+a_2
\mu +a_3 \mu^2),
$$
$$
P(\mu)=\sqrt{1-m_0/(m_0+\mu)^2},
$$
$$
\mu=\sqrt{s}-m_0,
$$
where the $a_i$ are free parameters, $P(\mu)$ is an approximation
of the two-body phase space for particles with similar masses.
Both the $\phi(1020)$ and $\fzero$ have small but finite widths,
and the selection criterion of $m_{\pp}>0.85$~$\hbox{GeV}/c^2$
defines an effective minimum mass, $m_0=1.8~\hbox{GeV}/c^2$. For
the signal resonance, we take the same form as described in $\phi
\pp$ mode, except that the three-body decay phase space factor is
replaced by the two-body decay phase space factor.

\begin{figure}[htbp]
\centerline{\psfig{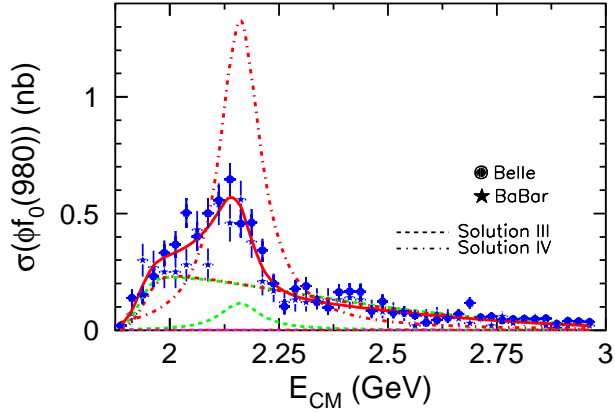} } \caption{ Fit to
$\EE\to \phi \fzero$ cross section with a single $BW$ function that
interferes with a nonresonant component. The curves show the
projections from the best fit and the contribution from each
component. The dashed curves are for constructive solution, and the
dot-dashed curves for destructive solution. } \label{phif0-two}
\end{figure}

We fit the $\phi \fzero$ cross section distribution with a single
$BW$ function that interferes with a nonresonant component.
Figure~\ref{phif0-two} shows the result. There are two solutions.
The interference is constructive for one solution (dashed curves)
and destructive for the other (dot-dashed curves). The values of the
fitted results for these two solutions are shown in
Table~\ref{phif0-twosolution}, with a goodness-of-the-fit of
$\chi^2/ndf=102/80$.

\begin{table}[htbp]
\caption{Fit results to the combined BaBar and Belle data on $\EE
\to \phi \fzero$ with a single $BW$ function that interferes with
a nonresonant component. The errors are statistical only. $M$,
$\Gamma_{{\rm tot}}$, and $\BR(\phi \fzero)\times \Gamma_{\EE}$
are the mass (in~MeV/$c^2$), total width (in~MeV/$c^2$), and
product of the branching fraction to $\phi \fzero$ and the $\EE$
partial width (in~eV/$c^2$), respectively. $\phi$ is the relative
phase (in degree).} \label{phif0-twosolution}
\begin{center}
\begin{tabular}{c  c  c }
\hline \hline
  Parameters & Solution I & Solution II \\\hline
 $M(\y)$ & \multicolumn{2}{c}{$2159\pm11$} \\
  $\Gamma_{{\rm tot}}(\y)$&   \multicolumn{2}{c}{$113\pm19$}\\
 $\BR(\phi \fzero)\times
\Gamma_{\EE}(\y)$ & $4.1\pm1.1$ & $47.6\pm1.5$\\
 $\phi$ & $130\pm12$ & $-106\pm2$ \\
 \hline \hline
\end{tabular}
\end{center}
\end{table}

\begin{figure*}[htbp]
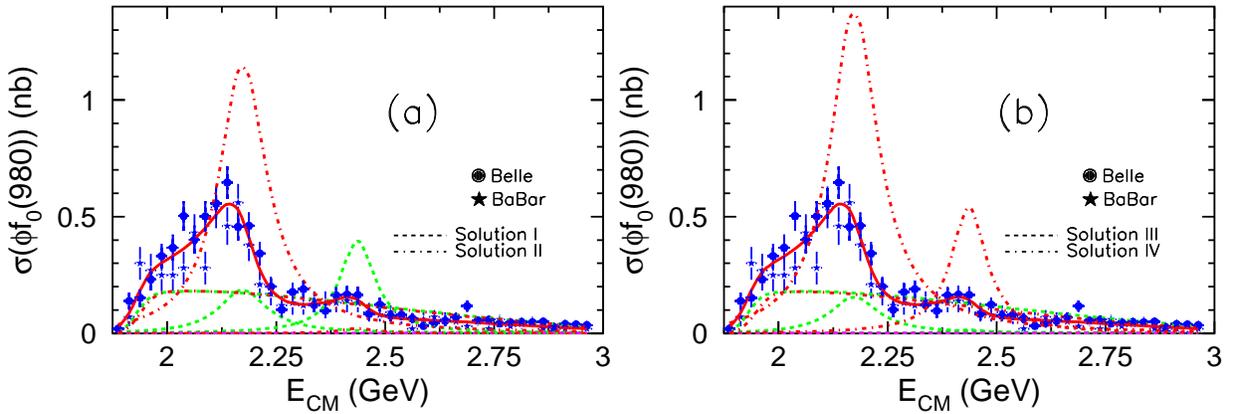

\centerline{\psfig{file=fig6a.epsi,width=8cm}
\psfig{file=fig6b.epsi,width=8cm} } \caption{Fit to $\EE\to \phi
\fzero$ cross sections with two coherent $BW$ functions that
interfere with a nonresonant component. The curves show the
projections from the best fit and the contribution from each
component. There are four solutions.} \label{phif0-three}
\end{figure*}

We also fit the $\phi \fzero$ cross section distribution with two
coherent $BW$ functions (one for the $\y$ and the other for the
structure at about 2.4~GeV), which interfere with a nonresonant
component. There are four solutions. Figures~\ref{phif0-three}(a)
and (b) show the result. The values of the fitted results for
these four solutions are shown in Table~\ref{phif0-foursolution},
with a goodness-of-the-fit of $\chi^2/ndf=88/76$. The fitted
values of the mass and width of the $\y$ are consistent with the
results obtained with a single $BW$ function interfering with a
nonresonant component within $1\sigma$, and the statistical
significance of the structure at $2.4~\hbox{GeV}/c^2$ is estimated
to be $2.7\sigma$.

\begin{table*}[htbp]
\caption{Fit results to the combined BaBar and Belle data on $\EE
\to \phi \fzero$ with two coherent $BW$ functions that interfere
with a nonresonant component. The errors are statistical only.
$M$, $\Gamma_{{\rm tot}}$, and $\BR(\phi \fzero)\times
\Gamma_{\EE}$ are the mass (in~MeV/$c^2$), total width
(in~MeV/$c^2$), and product of the branching fraction to $\phi
\fzero$ and the $\EE$ partial width (in~eV/$c^2$), respectively.
$\phi$ is the relative phase (in degree).}
\label{phif0-foursolution}
\begin{center}
\begin{tabular}{c  c  c c c}
\hline \hline
Parameters & Solution I & Solution II & Solution III &
Solution IV \\\hline
 $M(\y)$ & \multicolumn{4}{c}{$2171\pm16$} \\
  $\Gamma_{{\rm tot}}(\y)$&   \multicolumn{4}{c}{$134\pm22$}\\
 $\BR(\phi \fzero)\times
\Gamma_{\EE}(\y)$ & $8.0\pm3.3$ & $49.6\pm2.0$ & $6.7\pm1.3$ & $59.6\pm1.9$\\
 $\phi_1 $ & $172\pm38$ & $-116\pm3$ & $153\pm7$ & $-97\pm2$\\
 $M(X(2400))$ & \multicolumn{4}{c}{$2436\pm34$} \\
  $\Gamma_{{\rm tot}}(X(2400))$&   \multicolumn{4}{c}{$99\pm105$}\\
 $\BR(\phi \fzero)\times
\Gamma_{\EE}(X(2400))$ & $15.8\pm12.8$ & $0.8\pm0.4$ & $0.6\pm0.3$ & $21.5\pm1.3$\\
 $\phi_2$ & $-92\pm22$ & $91\pm21$ & $151\pm17$ & $-150\pm3$\\
 \hline \hline
\end{tabular}
\end{center}
\end{table*}

%%%%%%%%%%%%%%%%%%%%%%%%%%%%%%%%%%%%%%%%%%%%%%%%%%%%%%%%%%%
\section {Systematic errors}

The sources of the systematic errors on the resonance parameters
measurements are not very different from those listed by the Belle
experiment~\cite{belle-phipp}. For the masses and widths of the
$\phi(1680)$ and $\y$ resonances, we have considered the
uncertainties in the absolute mass scale, the mass resolution, the
background shape, fit range and possible existence of additional
resonances as systematic errors. For the products of the branching
fractions to $\phi \pp$ or $\phi \fzero$ and the $\EE$ partial
widths of the $\phi(1680)$ and $\y$ resonances, we have considered
the uncertainties in the background shape, fit range and cross
sections measurements as systematic errors. The sources of the
uncertainties in the cross sections measurements from Belle and
BaBar are independent, and the total systematic errors on the
cross sections measurements are 5.4\% and 4.9\% for $\phi \pp$ and
$\phi \fzero$ modes, respectively.

Finally, the results of the resonance parameters of the
$\phi(1680)$ and $\y$ are listed in Table~\ref{sum}, where the
first errors are statistical and the second systematic. The fitted
results with an additional $BW$ for the structure at around
2.4~GeV/$c^2$ are not included since the statistical significance
is smaller than 3.0$\sigma$.

\begin{table*}[htbp]
\caption{The results of the resonance parameters of the
$\phi(1680)$ and $\y$. The first errors are statistical and the
second systematic. $M$, $\Gamma_{{\rm tot}}$, and $\BR \times
\Gamma_{\EE}$ are the mass (in MeV/$c^2$), total width (in
MeV/$c^2$), and product of the corresponding branching fraction
and the $\EE$ partial width (in eV/$c^2$). There are two solutions
in fitting with coherent sum of the amplitudes.} \label{sum}
\begin{center}
\begin{tabular}{c  c c c c | c c}
\hline \hline State & Mode & Fit Method & Mass & $\Gamma_{{\rm
tot}}$ & \multicolumn{2}{c}{$\BR \times \Gamma_{\EE}$}
\\\hline
$\phi(1680)$ & $\phi \pp$ & incoherent & $1685\pm5\pm 3$ & $208\pm 11\pm 3$ & \multicolumn{2}{c}{$24.6\pm1.2\pm1.3$}\\
 &  $\phi \pp$ & coherent & $1677\pm6\pm 5$ & $233\pm 16\pm 15$ & $33.4\pm1.3\pm1.8$ &
 $23.0\pm1.5\pm1.3$\\\hline
$\y$ & $\phi \pp$ & incoherent & $2080\pm 12\pm 3$ & $182\pm 20\pm 10$ & \multicolumn{2}{c}{$18.1\pm1.8\pm 0.9$}\\
&  $\phi \pp$ & coherent & $2112\pm 16\pm 22$ & $196\pm 27\pm 25$ & $68.9\pm 7.0 \pm 3.4$ & $6.2\pm1.1\pm0.3$\\
& $\phi \fzero$ & coherent & $2159\pm 11\pm 13$ & $113\pm 19\pm 22$ & $47.6\pm 1.5 \pm 2.3$ & $4.1\pm1.1\pm0.4$\\
\hline \hline
\end{tabular}
\end{center}
\end{table*}

%%%%%%%%%%%%%%%%% summary %%%%%%%%%%%%%%%%%%%%%%%%%%%%%%%%%%%%%%%%%%%%%%%
\section {Results and discussion}

In summary, we present a fit to the cross sections for $\EE\to\phi
\pp$ and $\EE \to \phi \fzero$ from threshold to
$\sqrt{s}=3.0$~GeV measured by BaBar and Belle experiments. The
masses, widths and the products of the branching fraction and the
$\EE$ partial width of the $\phi(1680)$ and $\y$ are determined,
as listed in Table~\ref{sum}. From the table, we could see that
the differences in the $\phi(1680)$ resonance parameters between
two-incoherent-$BW$ fit and two-coherent-$BW$ fit to $\phi\pp$ are
not large. We take a simple average as the central value and
enlarge the errors to cover all the possibilities. However, the
differences in the $\y$ resonance parameters from different fit
are very large due to the assumption on the background shape and
multi-solution problem, especially for the value of the product of
the branching fraction and the $\EE$ partial width, we only give
$\y$ mass and width here. Finally, we obtain
$M(\phi(1680))=(1681^{+10}_{-12})$~MeV/$c^2$, $\Gamma_{\rm
tot}(\phi(1680))=(221^{+34}_{-24})$~MeV/$c^2$ and
$\BR(\phi(1680)\to\phi \pp)\times
\Gamma_{\EE}(\phi(1680))=27.0^{+8.6}_{-5.9}$~eV/$c^2$, and
$M(\y)=(2117^{+59}_{-49})$~MeV/$c^2$, $\Gamma_{\rm
tot}(\y)=(164^{+69}_{-80})$~MeV/$c^2$. Here the uncertainties
include both statistical and systematic errors.

We find that the central value of the $\y$ width is larger than
the BaBar measurement. However it is consistent with the predicted
values with the assumptions that the $\y$ is an $s\bar{s}g$
hybrid~\cite{ssg} or a $2^3D_1$ $s\bar{s}$ state~\cite{2D}. The
widths of the $\phi(1680)$ and $\y$ are quite similar and both are
at the 200~MeV/$c^2$ level. This may suggest that the $\y$ be an
excited $\phi$ state. Although there is faint evidence of the
structure at around 2.4~GeV/$c^2$ in $\EE \to \phi \pp$ and $\phi
\fzero$ in both BaBar and Belle data, the statistical significance
is smaller than 3.0$\sigma$. Larger data sample is necessary to
confirm it.

%%%%%%%%%%%%%%%%\acknowledgments %%%%%%%%%%%%%%%%%%%%%%%%%%%%%%%%%%%%%
\acknowledgments

This work is supported in part by the Department of Energy under
Contract No. DE-FG02-04ER41291 (U Hawaii), National Natural
Science Foundation of China under Contract Nos. 10825524 and
10935008, and Major State Basic Research Development Program
(2009CB825203).

%%%%%%%%%%%%%%%%%%%%%%%%%%%%%%%%%%%%%%%%%%%%%%%%%%%%%%%%%%%%%%%%%%%%%%


\begin{thebibliography}{**}

\bibitem{PDG} C.~Amsler {\em et al.} (Particle Data Group),
Phys. Lett. B {\bf 667}, 1 (2008).

\bibitem{babar2175}
  B.~Aubert  {\it et al.}  (BaBar Collaboration),
  Phys.\ Rev.\ D  {\bf 74}, 091103(R) (2006);
  {\it ibid.}  {\bf 76}, 012008 (2007). Data used in this paper are
  available at http://durpdg.dur.ac.uk/cgi-bin/hepdata/testreac/5242/RED

\bibitem{belle-phipp} C.~P.~Shen {\it et al.}  (Belle
Collaboration), Phys.\ Rev.\ D  {\bf 80}, 031101(R) (2009). Data
used in this paper are available at
http://durpdg.dur.ac.uk/cgi-bin/hepdata/testreac/5279/RED

\bibitem{babay4260} B.~Aubert {\em et al.} (BaBar Collaboration),
  Phys.\ Rev.\ Lett.\ {\bf 95}, 142001 (2005).

\bibitem{bn978} C.~Z.~Yuan {\it et al.} (Belle Collaboration),
 Phys.\ Rev.\ Lett.\ {\bf 99}, 182004 (2007).

\bibitem{ssg}
  Gui-Jun Ding and Mu-lin Yan,  Phys.\ Lett.\  B {\bf 650}, 390 (2007).

\bibitem{2D}
  Gui-Jun Ding and Mu-lin Yan,  Phys.\ Lett.\  B {\bf 657}, 49 (2007).

\bibitem{4s}
  Zhi-Gang Wang, Nucl. Phys. A {\bf 791}, 106 (2007).

\bibitem{chenhx}
  Hua-Xing Chen, Xiang Liu, Atsushi Hosaka and Shi-Lin Zhu,
  Phys.\ Rev.\ D  {\bf 78}, 034012 (2008).

\bibitem{dren} N. V. Drenska, R. Faccini and A. D. Polosa, Phys. Lett. B {\bf
669}, 160 (2008).

\bibitem{eberhard}
  E.~Klempt and  A.~Zaitsev, Phys.\ Rept. {\bf 454}, 1 (2007).

\bibitem{susana} S.~Coito, G.~Rupp and E.~van~Beveren,
arXiv:0909.0051.

\bibitem{27mev}
  A.~Martinez Torres, K.~P.~Khemchandani, L.~S.~Geng, M.~Napsuciale and E.~Oset,
  Phys.\ Rev.\ D  {\bf 78}, 074031 (2008).

\bibitem{alv} L.~Alvarez-Ruso, J.~A.~Oller and J.~M.~Alarcon, Phys.\ Rev.\ D  {\bf 80}, 054011 (2009).

\bibitem{3s}
  T.~Barnes, N. Black and P. R. Page,  Phys.\ Rev.\ D {\bf 68}, 054014 (2003).

\bibitem{zhusl}
  Shi-Lin Zhu, Int. J. Mod. Phys. E {\bf 17}, 283 (2008).

\bibitem{minuit} F.~James, MINUIT, CERN Program Library Writeup
Report No. D506.

\bibitem{babar1680}  B.~Aubert  {\it et al.}  (BaBar Collaboration),
Phys.\ Rev.\ D  {\bf 77}, 092002 (2008).

\bibitem{bes2175}
  M.~Ablikim {\it et al.}  (BES Collaboration),
  Phys.\ Rev.\ Lett.\ {\bf 100}, 102003 (2008).

\bibitem{DM1} F.~Mane {\it et al.}, Phys.\ Lett.\  B {\bf 112}, 178 (1982).

%%%%%%%%%%%%%%%%%%%%%%%%%%%%
\end{thebibliography}
\end{document}